\documentclass[twocolumn,showpacs,preprintnumbers,amsmath,amssymb]{revtex4}

\def\bi{\bigskip}
\def\be{\begin{equation}}
\def\en{\end{equation}}
\def\bq{\begin{eqnarray}}
\def\eq{\end{eqnarray}}
\def\bqa{\begin{eqnarray*}}
\def\eqa{\end{eqnarray*}}
\def\noi{\noindent}

\usepackage{graphicx} 
\usepackage{amsmath}
\usepackage{dcolumn}

\begin{document}

\title{The role of degeneracy in the analogy between \\continuous variable and spin $1/2$ systems}

\author{J.L.~Lucio M.}
\email{lucio@fisica.ugto.mx}
\affiliation{Instituto de F\'{\i}sica, Universidad de Guanajuato \\
Loma del Bosque \# 103, Lomas del Campestre, \\
37150 Le\'on, Guanajuato; M\'exico}

\date{\today}
\begin{abstract}
    \noi We point out limitations to the analogy between the
continuous variable and spin $1/2$ systems and show that the
maximal violation of Bell inequality is related to an infinite
degeneracy. We quantify non-maximal violation of the Bell-CHSH
inequality and comment potential experimental implications of our
work.

\end{abstract}

\maketitle

\section{I\lowercase{ntroduction}}
The generalization of Bell`s inequalities to quantum systems with
continuous variables (CV`s) has been formulated in terms of the
Wigner representation \cite{bana} or else using a map between the
CV`s and a spin 1/2 system \cite{zbc}. In fact it has been shown
\cite{jeong} that the pseudospin operators introduced in
\cite{zbc} are a limiting case of the observable introduced by
Gisin and Peres \cite{gisper}, establishing thus a bridge between
the finite and infinite dimensional (continuous) systems. The
relevance of such studies stem from the fact that the violation of
Bell`s inequalities are considered a measure of the nonlocality
and quantum nature of the system under consideration. Conditions
under which maximal violation is attained has also been discussed
in detail \cite{jeong}.  \bi

\noi The formulation of Bell inequalities \cite{ref} requires the
existence of an observable with eigenvalues $\pm 1$. Once the
observable is chosen the Bell operator is built and the inequality
is expressed in terms of the expectation value for a given state.
The state used to define the expectation value is considered the
source of entanglement and non-locality, and therefore of the
inequality's violation. The authors in \cite{zbc} considered the
Bell inequality due to Clauser, Horne, Shimony and Holt
\cite{bellchsh} using the two mode squeezed vacuum state and the so
called pseudospin operator, which are introduced to realize the
mapping between the continuous variables and the spin $1/2$. In this
way, the authors show that the Bell-CHSH inequality can be maximally
violated. It is worth remarking that the two mode squeezed vacuum
state is a regularized version of the state used by EPR in his
famous article \cite{epr}. In fact, the EPR state is obtained in the
limit of infinite squeezing, coincidentally the maximal violation of
the Bell CHSH is obtained also in that limit. On the other hand, it
has been argued that the map between CV`s and the pseudospin
operator opens the possibility that the CV`s systems may be
exploited to do quantum information tasks which could be robust
against photon losses.\bi

\noi This work is concerned with the generalization of the Bell's
inequalities based upon the mapping between the CV's and a two qubit
spin ½ system. Two points motivated our study. First, one should be
careful with such mapping since intuition tell us that the degrees
of freedom, or the information contained in the CV`s system should
be reflected some how in the finite dimensional system (pseudospin).
In fact, it turns out that the mapping necessarily lead to a strong
degeneration. Second, it is known \cite{BMR1} that maximal violation
is closely related to degeneration. The purpose of this paper is to
quantify the role of degeneration in the violation of the
inequalities, {\it i.e} to calculate the violation of the Bell-CHSH
inequality as a function of the degeneracy. There are two ways to
achieve our goal. The first, and more direct one, is based upon the
use of different truncated versions (in the number representation
basis) of the two mode squeezed state. The second approach, the  one
we follow here, use the concept of "entangled observable"
\cite{rusos-tv}. The point is that usually the state used to define
the expectation value is considered the source of violation of the
inequality. However there exist also the possibility that are the
observables - and not the states - which have the characteristic of
being non-local and or entangled. \bi

\noi The possibility to transfer the entanglement from the state
to the observable is a consequence of the fact that the two mode
squeezed vacuum state can be obtained through an unitary (but
non-local) transformation from the vacuum and that, as for any
other unitary transformation, physics is the same in both cases.
Assigning the entanglement to the states or to the observable are
just two different ways to evaluate the matrix elements (although
each procedure has its own interpretation),  the advantage of the
approach we use here is that it permits to gain insight into the
role played by degeneracy. Section II is devoted to a brief
revision of the conventional approach to the inequality, section
III contains the contribution of this work and make a brief
comment regarding the potential experimental implications of our
work, finally in section IV we summarize our results.

\section{S\lowercase{pin 1/2 analogy}}

The Bell-CHSH inequality for a two qubit system is expressed in
terms of the Bell operator $\mathcal{B}$:

\begin{eqnarray} \label{chsh}
\mathcal{B}=(\mathbf{u}\cdot \hat{s}_1) \otimes
(\mathbf{v\cdot\hat{s}_2})
 + (\mathbf{u}\cdot\hat{s}_1)\otimes(\mathbf{v}'\cdot\hat{s}_2)+
 \nonumber
 \\
 (\mathbf{u}'\cdot \hat{s}_1) \otimes (\mathbf{v\cdot\hat{s}_2}) -
 (\mathbf{u}'\cdot\hat{s}_1)\otimes(\mathbf{v}'\cdot\hat{s}_2).
\end{eqnarray}

\noi where $\hat{s}_i$ is the Pauli matrix for the i-th qubit $(i =
1,2)$ and $\textbf{u},\textbf{v},\textbf{u}'$ and $\textbf{v}'$, are
three dimensional unit vectors. For a local realistic theory the
Bell-CHSH inequality holds \cite{bellchsh} $|\mathcal{<B>}| \leq 2$
while for any two qubit state Cirel'son limit applies
$|\mathcal{<B>}| \leq 2\sqrt{2}$. \bi

\noi Two mode squeezed vacuum states (TMSV) are relevant in the
generalization of the Bell-CHSH inequality (\ref{chsh}) for
continuous variables (CV) systems since matrix elements are taken
respect to these states. TMSV are constructed by means of the
creation and annihilation operators $(a,a^\dagger)$ and
$(b,b^\dagger)$ associated to the first and second channels
respectively:

\bq \label{tmsv}  S(\zeta) &=&
e^{\zeta(a^\dagger b^\dagger - a b)}\nonumber\\
|\zeta>=S(\zeta)|00> &=&
\sum_{n=0}^{\infty}\frac{(\tanh(\zeta))^n}{\cosh(\zeta)}|nn> \eq

\noi where $\zeta > 0$ is the squeezing parameter and $|nn>
\hspace{.15cm} \equiv |n_a> \otimes \hspace{.15cm} |n_b> $. In
analogy with the harmonic oscillator it proofs convenient to
introduce "position and momentum operators" through the relation $(a
= (\textbf{q}+i\textbf{p})/\sqrt{2})$ and similarly for the second
channel. As usual, eigenstates of the number operator $N_a =
a^\dagger a$ ( $N_b = b^\dagger b$) are denoted by $|n> $. For a
single mode light field, the authors in \cite{zbc} introduce the
pseudospin operators for photons:

\bq \label{pspin}
 s_z &=& \sum^\infty_{n=0}[|2n+1 \rangle\langle 2n+1|-|2n\rangle
 \langle 2n|] \nonumber \\
 s_-&=&\sum^\infty_{n=0} |2n \rangle\langle 2n+1|  \\
 s_+ &=& (s_-)^\dagger,\hspace{0.5cm} s_\pm=\frac{1}{2}(s_x\pm
 is_y) \nonumber
\eq

\noi For an arbitrary unit vector $\textbf{u}$ defined by the
polar $(\theta_a)$ and azimuthal $(\phi_a)$ angles, the projection
of the spin operator on the direction of $\textbf{u}$ is given by:

\begin{equation}
\mathbf{u}\cdot\hat{s} = s_z \cos \theta_u + \sin \theta _u
 (e^{i\varphi_u} s_- + e^{-i\varphi_u} s_+) \nonumber
\end{equation}

\noi Since the commutation relation for the operators $s_z,s_-$
and $s_+$ are identical to those of the spin $1/2$ system and
given that for arbitrary unit vector $\textbf{u}$,
$(\textbf{u}\cdot \hat s)^2 = 1$ the authors in \cite{zbc}
conclude that there exist a perfect analogy between CV and the
usual spin 1/2 systems. An alternative approach \cite{gkr} based
of the eigenstate $|q >$ of the position operator is available. In
this case the parity operator is introduced:

\begin{equation} \label{parity}
\mathbf{\Pi}_z = \int^\infty_0 \mathrm{dq}(|\varepsilon
\rangle\langle \epsilon | -
  |\mathcal{O}\rangle\langle\mathcal{O} |)
\end{equation}

\noi while the $x$ and $y$ components are chosen as:

\bq\label{pixy}
\mathbf{\Pi}_x =\int^\infty_0
\mathrm{dq}(|\epsilon \rangle\langle \mathcal{O} | +
  |\mathcal{O}\rangle\langle\varepsilon |) \\
\mathbf{\Pi}_y = i \int^\infty_0 \mathrm{dq}(|\mathcal{O}
\rangle\langle \epsilon | -
  |\varepsilon\rangle\langle\mathcal{O} |) \eq

\noi where the even and odd eigenstates are related to the
position eigenkets by the relations:

\bq  \label{qbasis} |\mathcal{E}> = \frac{1}{\sqrt{2}}(|q> + |-q>)
\nonumber \\ |\mathcal{O}> = \frac{1}{\sqrt{2}}(|q> - |-q>). \eq

\noi The components of $\mathbf{\Pi}$ also satisfy the $SU(2)$
algebra.
\bi

\noi For the CV case, the Bell operator is obtained by replacing
in (\ref{chsh}) the spin $1/2$ operators either by the pseudospin
(\ref{pspin}) or by the $\Pi$ operator in
(\ref{parity},\ref{pixy}). The inequality is expressed in terms of
the matrix elements of the Bell operator between TMSV states
$|\zeta>$. \bi

\noi The Bell operator (\ref{chsh}) involves four unit vectors.
The following values for the angles are chosen:
$\phi_u=\phi_v=\phi_u'=\phi_v'=0$, so that the matrix element
reduces to:

\begin{equation}  \label{reschshp}
<\zeta|\mathcal{B}|\zeta> = \cos(\theta_u)\cos(\theta_v)I(\zeta)+
\sin(\theta_u)\sin(\theta_v)F(\zeta)
\end{equation}

\noi where

\begin{eqnarray}  \label{IF}
I(\zeta) &=& <\zeta|s^1_z\otimes s^2_z|\zeta>, \nonumber \\
F(\zeta) &=& <\zeta|s^1_x\otimes s^2_x|\zeta>
\end{eqnarray}

\noi If we further choose $\theta_u=0$, $\theta_u'=\pi/2$,
$\theta_v=-\theta_v'$, we obtain \cite{zbc,gkr}:

\begin{equation}  \label{dachsh}
<\zeta|\mathcal{B}|\zeta> = 2 \cos(\theta_v)I(\zeta)+ 2
\sin(\theta_v)F(\zeta)
\end{equation}

\noi Maximal violation of the inequality is obtained for
$\tan(\theta_v)= F(\zeta)/I(\zeta)$ and amounts to:

\begin{equation}  \label{biqv}
<\zeta|\mathcal{B}|\zeta> = 2 \sqrt{I(\zeta)^2+ F(\zeta)^2 }
\end{equation}

\noi For the TMSV state $|\zeta>$, $I(\zeta)=1$ whereas $F(\zeta)$
depends upon our choice for the $x,y$ components of the pseudospin
operators. Thus for example \cite{zbc,gkr}:

\bq
F(\zeta)&=& \tanh(2\zeta) \hspace{0.25cm} for \hspace{0.15cm} (\ref{pspin}), \\
F(\zeta)&=& \frac{2}{\pi} \arctan(\sinh(2\zeta)) \hspace{0.20cm}
 for \hspace{0.10cm} (\ref{pixy}).
\eq

\noi Note that in both cases the Cirel'son bound is attained in
the $\zeta \to \infty$ limit, although for all $\zeta$,
$\tanh(2\zeta) \leq \frac{2}{\pi}\arctan(\sinh(2\zeta))$.

\section{T\lowercase{he role of degeneracy}}

The analogy between the CV system and the spin $1/2$ is appealing,
however it is clear that care must be exerted since, in general we
do not expect both systems to have similar properties . For
example, the number of degrees of freedom involved in both systems
are not equal. An important difference between the conventional
spin $1/2$ and the pseudospin operators introduced in \cite{zbc}
is the degeneracy. For the spin $1/2$ there is a unique $|+>$
state such that $s_z|+>= \frac{\hslash}{2}|+>$ while for the
pseudospin operator (\ref{pspin}) all the states of the type
$|2n_0+1>$ for $n_0=0,1,2, ...\infty$ are eigenstates of the
parity-spin with eigenvalue one. Thus, there is an infinite
degeneracy. Similarly for spin $1/2$ there is only one state such
that $s_+|+> = 0$ while for pseudospin all the states $|2n_0+1>$
with $n_0=0,1,2, ..., \infty$ are annihilated by $s_+$. Similar
results hold when eigenstates $|q>$ of the position operator are
considered since in this case any symmetric (or antisymmetric)
$|q_0>$ state are eigenstates of the parity operator. \bi

\noi When considering the violation of the Bell-CHSH inequality for
squeezed states, an intriguing possibility  is the alternative of
redefining the operator. Indeed, instead of considering the matrix
element between TMSV states, we can consider expectation values
respect to the vacuum:

\begin{equation}  \label{reschsh}
<\zeta|\mathcal{B}|\zeta> = <00|\mathcal{\tilde{B}}|00>,
\end{equation}

\noi where obviously

\begin{equation}  \label{reschsh}
\mathcal{\tilde{B}}= S(\zeta)^\dagger \mathcal{B} S(\zeta)
\end{equation}

\noi Thus, instead of entangled states we have an entangled
observable \cite{rusos-tv}. In this example in particular, the
transformation $S(\zeta)$ is unitary and the non triviality of its
action stems from the fact that it involves operators acting on
different Hilbert spaces.\bi

\noi Thus, in order to evaluate the violation of the inequality,
we first  calculate the "entangled Bell CHSH" operator
(\ref{reschsh}) and then take its vacuum expectation value. Since
such an evaluation may become involved, we simplify the
calculation considering different approximations. To this end
notice that the analogy between the CV and the spin $1/2$ does not
require to keep an infinite number of terms in the definition of
the pseudospin operators in (\ref{pspin}). Thus for example,
restraining the sum to the $N=0$ term, the resulting operators
have the same properties in the corresponding subspace, as the
full pseudospin operator. We can work with those limited operators
so as to figure out the answer to the questions risen above. \bi

\noi Below we quote the expressions for the tensor product of the
pseudospin components entering in $\mathcal{\tilde{B(\zeta)}}$
(see (\ref{reschsh}, \ref{reschshp}, \ref{IF})) when the sum in
(\ref{pspin} )is limited to the first term:

\bq &&\cosh(\zeta)^2s_z^{(1)}\otimes s_z^{(2)} \to e^{K
ab}s_z^{(1)} \otimes s_z^{(2)} e^{K a^\dagger b^\dagger}
\\&=&(1+K^2)|00><00|-|01><01|-|10><10| \nonumber  \\
&+&|11><11| + K (|11><00|+ |00><11|) \nonumber  \eq

\bq &&\cosh(\zeta)^2s_x^{(1)}\otimes s_x^{(2)} \to e^{K
ab}s_x^{(1)} \otimes s_x^{(2)} e^{K a^\dagger b^\dagger}
\\&=&|00><11|-|01><10|-|10><01| \nonumber  \\
&+&|11><00| + 2K |00><00|. \nonumber  \eq

\noi It should be clear that in the case we are considering there is
no degeneration. Since

\bq \label{nodeg} I(\zeta)_{N=0}= \frac{1+K^2}{(\cosh\zeta)^2},
\hspace{0.5cm} F(\zeta)_{N=0}= \frac{2K}{(\cosh\zeta)^2} \eq

\noi the Bell-CHSH inequality yields the following result:

\bq \label{nodege}
 <00|\mathcal{\tilde{B(\zeta)}}|00>_ {N=0}=\frac{1+6K^2+K^4}{(\cosh \zeta)^4}
\eq

\begin{figure}
 \includegraphics[width=8cm]{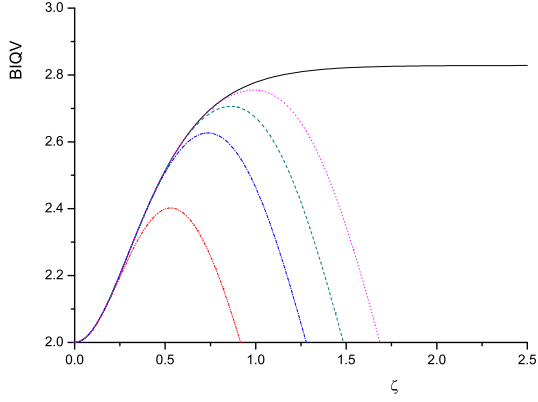}\\
  \caption{Bell inequality violation (BIQV) as a function of
  the squeezing parameter $\zeta$. The continuous line corresponds
  to $\infty$ degeneration. The other lines correspond to degeneracy
  equal to 3,2,1 and 0, in that order, from top to bottom. }
\end{figure}

\noi So far we know the vacuum expectation value of the Bell
operator when there is no degeneracy (\ref{nodeg},\ref{nodege}),
other cases may be treated along similar lines {\it i.e.}
increasing the degree of degeneration or equivalently increasing
number of terms one considers in the definition of the pseudospin
operator \footnote{The number of terms to be included for each
level of degeneration is fixed so that the commutation relations
are fulfilled. Due to the form of $s_{+}$ ($|2n_0+1><2n_0|$), the
minimal elements necessary to define a spin operator are two.
Correspondingly the number of terms required for $s_z$ are two
($|2n_0+1><2n_0+1| - |2n_0><2n_0|$). Thus, every time we increase
in one unit the level of degeneration we must add one term to
$s_{\pm}$ and two to $s_z$ }. Although the evaluation of the
tensor product $e^{K ab} s_i^{(1)}\otimes s_i^{(2)}e^{K a^\dagger
b^\dagger}$ very rapidly becomes cumbersome, the calculation is
simplified remembering that we only need the vacuum expectation
value of these operators. The expansion of the operator ($e^{K
ab}$) involves equal powers of the operators $a$ and $b$, so that
only terms of equal occupation number will survive the matrix
elements, so we write for the tensor products:

\bq \label{transs} s_z^{(1)}\otimes s_z^{(2)}&=&
\sum_{n=0}^i|n,n><n, n| +  h.c.\\
 s_x^{(1)}\otimes s_x^{(2)}&=& \sum_{n=0}^i|2n, 2n><2n+1, 2n+1| + h.c. \nonumber
\eq

\noi the upper limit indicates that we can stop the sum at the
$i-th$ term, in whose case we will have degeneracy equal to $i$,
$i=0$ corresponding to no degeneration. The action of the
operators $e^{K ab}$ to the left and $e^{K a^\dagger b^\dagger}$
to the right is readily calculated by expanding them in power
series, thus we obtain

\bq \label{transs}
 e^{K ab} s_z^{(1)}&\otimes& s_z^{(2)}e^{K a^\dagger b^\dagger}=
\sum_{n=0}^i (1+K^2)K^{4n}\noi<00|+ T_1 \nonumber\\
 e^{K ab} s_i^{(1)}&\otimes& s_i^{(2)}e^{K a^\dagger b^\dagger}
= 2\sum_{n=0}^i K^{4n+1}|00><00|+T_2.\nonumber\\
\eq

\noi where $T_1$ and $T_2$ stand for other terms whose vacuum
expectation value vanishes. This is to be compared to the
conventional calculation, where the $|\zeta>$ state is explicitly
introduced:

\bq  \label{convenx}
I(\zeta)&=&\sum_{n=0}^{\infty}\sum_{m=0}^{\infty}
\frac{K^{n+m}}{\cosh(\zeta)^2}<mm|s_x^{(1)}\otimes
s_x^{(2)}|nn> \nonumber \\
&=&2 \sum_{j=0}^{\infty}\frac{K^{4j+1}}{\cosh(\zeta)^2}
=\tanh(2\zeta) \eq

\bq  \label{convenz}
F(\zeta)&=&\sum_{n=0}^{\infty}\sum_{m=0}^{\infty}
\frac{K^{n+m}}{\cosh(\zeta)^2}<mm|s_z^{(1)}\otimes
s_z^{(2)}|nn> \nonumber \\
&=& \sum_{j=0}^{\infty}\frac{K^{2j}}{\cosh(\zeta)^2}=1. \eq

\bi

\begin{figure}
 \includegraphics[width=8cm]{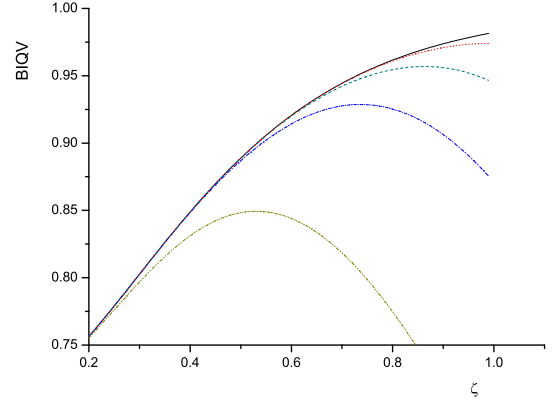}\\
  \caption{Ratio of the Bell inequality violation (BIQV)to the
  maximal possible value $(2\sqrt{2})$ as a function of the squeezing
parameter $\zeta$. The continuous line corresponds
  to $\infty$ degeneration. The other lines correspond to degeneracy
  equal to 3,2,1 and 0, in that order, from top to bottom. }
\end{figure}

\noi (\ref{transs}) and (\ref{convenx},\ref{convenz}) lead to the
same result when an infinite number of terms are included. The
advantage of (\ref{transs}) is that it permits to quantify the
role of degeneracy in the violation of Bell's inequality. In
Fig(1) we show the vacuum expectation value of the Bell operator,
as a function of $\zeta$, for different levels of degeneration .
In particular, the results for infinite degeneration ($i=\infty$),
obtained by Chen {\it et al} \cite{zbc} and the result when no
degeneration ($i=0$) is present (\ref{nodege}) are shown. Notice
that the behavior for small values of $\zeta$ is as interesting as
the $\zeta \to \infty$ region. Indeed, for the two mode squeezed
vacuum states (TMSV) and for $\zeta \to 0$, we know the violation
tend to vanish. In such a limit all the approximations (different
level of degeneration) coincide with the exact result. However,
for $1/2 < \zeta < 1$ we observe an important violation of the
Bell-CHSH inequality.  An alternative way to present this
information is by plotting (see Fig(2) ) the ratio of the
expectation value of the Bell operator for different level of
degeneration to the maximal possible value ($2 \sqrt{2}$). Clearly
the larger the degeneration the closer (and always below) the
vacuum expectation value to the behavior obtained for infinite
degeneracy. \bi

\noi  Maximal violation of Bell`s inequalities is predicted only
for infinite squeezing, which is not experimentally easy to
realize. In this respect note that for $\zeta \approx 1/2 $ and no
degeneration (only the $|00>, |01>, |10>$ and  $|11> $ states are
involved), the violation of Bell`s inequality may be as large as
$85\%$ of the Cirel'son bound. This may be relevant when
considering the robustness against photon losses of continuous
variable systems.

\section{S\lowercase{ummary}}
In summary, in this paper we:

\begin{itemize}
\item Qualified the statement regarding the "perfect analogy"
between a system with continuous variables and a spin $1/2$, by
remarking the existence of an infinite degeneration associated to
the pseudospin operators introduced by authors in \cite{zbc}.
\item  Worked out an example that permit us to introduce the
concept of entangled operator and also allow us to quantify the
degree of violation of Bell's inequalities as a function of the
degeneration and the squeeze parameter. \item Found a large
$(>85\%)$ violation of the Bell-CHSH inequality in the
non-degenerate case. For infinite degeneracy maximal violation is
recovered.
\end{itemize}

\begin{acknowledgments}
The authors acknowledge financial support from  CONACyT under
project 44644-F and CONCyTEG.
\end{acknowledgments}


\begin{thebibliography}{1}
\bibitem{bana} K. Banaziek and K. Wodkiewicz, Phys. Rev. A 58 (1998) 4345; Phys. Rev. Letts. 82 (1999) 2009.

\bibitem {zbc} Zeng-Bing Chen, Jian-Wei Pan, Guang Hou and Yong-De
Zhang, Phys. Rev. Letts. 88 (2002) 04046; L. Mysta, Jr., R. Filip
and J. Fiurasek, Phys. Rev. A65 (2002) 062315. See also C.
Brukner, M.S. Kim, Jian-Wei  Pan and A. Zeilinger, Phys. Rev A68
(2003) 062105.

\bibitem{jeong} H. Jeong, W. Son, M.S. Kim, D. Ahn and C. Bruckner,
Phys. Rev. A67 (2003) 012106.

\bibitem{gisper} N. Gisin and A. Peres, Phys. Letts. A 162 (1992) 15.

\bibitem{ref} A detailed account of the efforts to describe
non-locality and entanglement both for discrete and continuous
variables can be found in the review article by S.L. Braunstein,
and P. Loock, Reviews of Modern Physics 77 (2005) 513.

\bibitem {bellchsh} J.S. Bell, \emph{Speakable and Unspeakable in Quantum Mechanics},
Cambridge University Press (1987) 139.

\bibitem{epr} A. Einstein, B. Podolsky and N. Rosen, Phys. Rev. 47 (1935) 777.

\bibitem{BMR1} S. L. Braunstein, A. Mann and M. Revsen, Phys. Rev.
Letts. 68 (1992) 3259.

\bibitem{rusos-tv} A.S. Trushechkim and I.V. Volovich, ArXiv:
quant-ph/0504156.

\bibitem{gkr} G. Gour, F.C. Khana, A. Mann, M. Revzen, Phys.
Letts. A 324 (2004) 415.
\end{thebibliography}
\end{document}